\documentclass[iop,apj,twocolappendix]{emulateapj}
\usepackage{apjfonts}

\usepackage[colorlinks]{hyperref}
\usepackage{amsmath,amsthm}
\usepackage{stmaryrd} 
\usepackage{upgreek} 
\usepackage{multirow}

\newtheorem{thm}{Theorem}

\newcommand{\NS}{NS}
\newcommand{\mbta}{MBTA}
\newcommand{\GW}{GW}
\newcommand{\EM}{EM}
\newcommand{\GRB}{GRB}
\newcommand{\CBC}{CBC}
\newcommand{\LIGO}{LIGO}

\newcommand{\LSO}{LSO}
\newcommand{\SNR}{S/N}
\newcommand{\Msun}{\ensuremath{M_{\odot}}}

\newcommand{\tmpsamps}{\ensuremath{N}}
\newcommand{\numtmps}{\ensuremath{M}}
\newcommand{\numslices}{\ensuremath{S}}
\newcommand{\SVD}{SVD}
\newcommand{\svdtmps}[1]{\ensuremath{L^#1}}
\newcommand{\numsvdtmps}{\svdtmps{s}}
\newcommand{\slicesamps}[1]{\ensuremath{N^#1}}
\newcommand{\slicessamps}{\slicesamps{s}}
\newcommand{\fftblock}{\ensuremath{D}}

\newcommand{\fir}{FIR}
\newcommand{\fft}{FFT}

\newcommand{\flops}{flop~s$^{-1}$}
\newcommand{\gstlal}{{\tt gstlal}}
\newcommand{\gstreamer}{GStreamer}

\newcommand{\lloid}{LLOID}
\newcommand{\TD}{TD}
\newcommand{\FD}{FD}

\DeclareMathOperator{\expectation}{E}
\DeclareMathOperator{\var}{var}
\DeclareMathOperator{\sinc}{sinc}

\def\clap#1{\hbox to 0pt{\hss#1\hss}}

\def\mathclap{\mathpalette\mathclapinternal}

\def\mathclapinternal#1#2{\clap{$\mathsurround=0pt#1{#2}$}}

\begin{document}

\title{Toward early-warning detection of gravitational waves from compact binary coalescence}

\slugcomment{Received 2011 August 16; accepted 2012 January 25; published 2012 March 15}

\shorttitle{Early-warning detection of \GW\ inspirals}

\author{
	Kipp Cannon\altaffilmark{1},
	Romain Cariou\altaffilmark{2},
	Adrian Chapman\altaffilmark{3},
	Mireia Crispin-Ortuzar\altaffilmark{4},
	Nickolas Fotopoulos\altaffilmark{3},
	Melissa Frei\altaffilmark{5,6},
	Chad Hanna\altaffilmark{7},
	Erin Kara\altaffilmark{8},
	Drew Keppel\altaffilmark{9,10},
	Laura Liao\altaffilmark{11},
	Stephen Privitera\altaffilmark{3},
	Antony Searle\altaffilmark{3},
	Leo Singer\altaffilmark{3}, and
	Alan Weinstein\altaffilmark{3}
}

\altaffiltext{1}{Canadian Institute for Theoretical Astrophysics, Toronto, ON, Canada}
\altaffiltext{2}{D\'{e}partement de physique, \'{E}cole Normale Sup\'{e}rieure de Cachan, Cachan, France}
\altaffiltext{3}{LIGO Laboratory, California Institute of Technology, MC 100-36, 1200 E. California Blvd., Pasadena, CA, USA; \texttt{leo.singer@ligo.org}}
\altaffiltext{4}{Facultat de F\'{i}sica, Universitat de Val\`{e}ncia, Burjassot, Spain} 
\altaffiltext{5}{Department of Physics, University of Texas at Austin, Austin, TX, USA}
\altaffiltext{6}{Center for Computational Relativity and Gravitation and School of Mathematical Sciences, Rochester Institute of Technology, Rochester, NY, USA}
\altaffiltext{7}{Perimeter Institute for Theoretical Physics, Waterloo, ON, Canada}
\altaffiltext{8}{Department of Physics and Astronomy, Barnard College, Columbia University, New York, NY, USA}
\altaffiltext{9}{Albert-Einstein-Institut, Max-Planck-Institut f\"{u}r Gravitationphysik, Hannover, Germany}
\altaffiltext{10}{Institut f\"{u}r Gravitationsphysik, Leibniz Universit\"{a}t Hannover, Hannover, Germany}
\altaffiltext{11}{Department of Chemistry and Biology, Ryerson University, Toronto, ON, Canada}


\begin{abstract}
Rapid detection of compact binary coalescence (CBC) with a network of advanced
gravitational-wave detectors will offer a unique opportunity for
multi-messenger astronomy.  Prompt detection alerts for the astronomical
community might make it possible to observe the onset of electromagnetic
emission from CBC.  We demonstrate a computationally
practical filtering strategy that could produce early-warning triggers before
gravitational radiation from the final merger has arrived at the detectors.
\end{abstract}

\section{Introduction}

As a compact binary system loses energy to gravitational waves (\GW{}s), its
orbital separation decays, leading to a runaway inspiral with the \GW\
amplitude and frequency increasing until the system eventually merges.  If a
neutron star (\NS) is involved, it might become tidally disrupted near the
merger and fuel an electromagnetic (\EM) counterpart \citep{shibata:2007}.
Effort from both the \GW\ and the broader astronomical communities might make
it possible to use \GW\ observations as early warning triggers for \EM\
follow-up. In the first generation of ground-based laser interferometers, the
\GW\ community initiated a project to send alerts when potential \GW\
transients were observed in order to trigger follow-up observations by \EM\
telescopes.  The typical latencies were 30 minutes \citep{HugheyGWPAW2011}.
This was an important achievement, but too late to catch any prompt optical flash
or the onset of an on-axis optical afterglow.  Since the \GW\ signal is in principle detectable even before the tidal
disruption, one might have the ambition of reporting \GW\ candidates not minutes
after the merger, but seconds before.  We explore one essential ingredient of
this problem, a computationally inexpensive latency free, real-time filtering
algorithm for detecting inspiral signals in \GW\ data.  We also consider the
prospects for advanced \GW\ detectors and discuss other areas of work that would
be required for rapid analysis.

Compact binary coalescence (\CBC) is a plausible progenitor for most short
gamma-ray bursts \citep[short \GRB{}s;][]{Lee:2005, nakar07}, but the
association is not iron-clad \citep{2011ApJ...727..109V}. The tidally
disrupted material falls onto the newly formed, rapidly spinning compact object
and is accelerated in jets along the spin axis with a timescale of $0.1$--$1$~s
after the merger \citep{Janka1999}, matching the short \GRB\ duration
distribution well. Prompt \EM\ emission including the \GRB\ can arise as fast
outflowing matter collides with slower matter ejected earlier in inner shocks.
The same inner shocks, or potentially reverse shocks, can produce an
accompanying optical flash \citep{Sari99}. The prompt emission is a probe into
the extreme initial conditions of the outflow, in contrast with afterglows,
which arise in the external shock with the local medium and are relatively
insensitive to initial conditions. Optical flashes have been observed for a
handful of long \GRB{}s \citep{2011CRPhy..12..255A} by telescopes with extremely
rapid response or, in the case of GRB 080319b, by pure serendipity, where
several telescopes were already observing the afterglow of another \GRB\ in
the same field of view \citep[FOV;][]{2008Natur.455..183R}. The observed optical flashes
peaked within tens of seconds and decayed quickly. For short \GRB\ energy
balance and plasma density, however, the reverse shock model predicts a peak
flux in radio, approximately 20 minutes after the \GRB, but also a relatively
faint optical flash \citep{nakar07}; for a once-per-year Advanced LIGO event at
$130$~Mpc, the radio flux will peak around 9~GHz at $\sim$$5$~mJy, with emission
in the $R$-band at $\sim$19~mag. Interestingly, roughly a quarter to half of the 
observed short \GRB{}s also exhibit extended X-ray emission of $30$--$100$~s in
duration beginning $\sim$$10$~s after the \GRB\ and carrying comparable fluence
to the initial outburst. This can be explained if the merger results in the
formation of a proto-magnetar that interacts with
ejecta~\citep{Bucciantini2011}. Rapid \GW\ alerts would enable joint \EM\ and
\GW\ observations to confirm the short \GRB-\CBC\ link and allow the early
\EM\ observation of exceptionally nearby and thus bright events.

In 2010 October, \LIGO{}\footnote{\url{http://www.ligo.org/}} completed its
sixth science run (S6) and Virgo\footnote{\url{http://www.ego-gw.it/}}
completed its third science run (VSR3).  While both \LIGO\ detectors and Virgo
were operating, several all-sky detection pipelines operated in a low-latency
configuration to send astronomical alerts, namely, Coherent WaveBurst (cWB), Omega, and Multi-Band Template Analysis \citep[MBTA;][]{HugheyGWPAW2011, S6lowlatency2, S6lowlatency3, S6lowlatency4}.  cWB and Omega are both unmodeled searches for bursts based on time-frequency decomposition of the GW data.  \mbta\ is a novel kind of template-based inspiral search that was purpose-built for low latency operation.  \mbta\ achieved the best \GW\ trigger-generation latencies of 2--5 minutes.
Alerts were sent with latencies of 30--60 minutes, dominated by human vetting.
Candidates were sent for \EM\ follow-up to several telescopes; \textit{Swift},
LOFAR, ROTSE, TAROT, QUEST, SkyMapper,
Liverpool Telescope, Pi of the Sky, Zadko, and Palomar Transient Factory
\citep{kanner2008, HugheyGWPAW2011} imaged some of the most likely sky
locations.

There were a number of sources of latency associated with the search for
\CBC\ signals in S6/VSR3 \citep{HugheyGWPAW2011}, listed here.

\paragraph{Data acquisition and aggregation ($\gtrsim$100~ms)}%
The \LIGO\ data acquisition system collects data from detector subsystems 16
times a second~\citep{Bork2001}. Data are also copied from all of the \GW\
observatories to the analysis clusters over the Internet, which is capable of
high bandwidth but only modest latency.  Together, these introduce a latency of
$\gtrsim$$100$~ms.  These technical sources of latency could be reduced with
significant engineering and capital investments, but they are minor compared
to any of the other sources of latency.

\paragraph{Data conditioning ($\sim$1~min)}%
Science data must be calibrated using the detector's frequency response to
gravitational radiation.  Currently, data are calibrated in blocks of 16~s.
Within $\sim$1~minute, data quality is assessed in order to create veto flags.
These are both technical sources of latency that might be addressed with
improved calibration and data quality software for advanced detectors.

\paragraph{Trigger generation (2--5~min)}%
Low-latency data analysis pipelines
deployed in S6/VSR3 achieved an impressive latency of minutes.  However, second
to the human vetting process, this dominated the latency of the entire \EM\
follow-up process.  Even if no other sources of latency existed, this trigger
generation latency is too long to catch prompt or even extended emission.
Low-latency trigger generation will become more challenging with advanced
detectors because inspiral signals will stay in band up to 10 times longer.  In
this work, we will focus on reducing this source of latency.

\paragraph{Alert generation (2--3~min)}%
S6/VSR3 saw the introduction of low-latency astronomical alerts, which required
gathering event parameters and sky localization from the various online
analyses, downselecting the events, and calculating telescope pointings.  If
other sources of latency improve, the technical latency associated with this
infrastructure could dominate, so work should be done to improve it.

\paragraph{Human validation (10--20~min)}%
Because the new alert system was commissioned during S6/VSR3, all alerts were
subjected to quality control checks by human operators before they were
disseminated. This was by far the largest source of latency during S6/VSR3.
Hopefully, confidence in the system will grow to the point where no human
intervention is necessary before alerts are sent, so we give it no further
consideration here.

\paragraph{}

This work will focus on reducing the latency of trigger production.  Data
analysis strategies for advance detection of \CBC{}s will have to strike a
balance between latency and throughput. \CBC\ searches consist of banks of
matched filters, or cross-correlations between the data stream and a bank of
nominal ``template'' signals.  There are many different implementations of
matched filters, but most have high throughput at the cost of high latency, or
low latency at the cost of low throughput.  The former are epitomized by the
overlap-save algorithm for frequency-domain (\FD) convolution, currently the
preferred method in \GW\ searches.  The most obvious example of the latter is
direct time domain (\TD) convolution, which is latency-free.  However, its cost in floating
point operations per second is linear in the length of the templates, so it is
prohibitively expensive for long templates.  The computational challenges of low-latency \CBC\ searches are still more daunting for advanced detectors for which the inspiral signal remains in band for a large fraction of an hour~(see the Appendix).

Fortunately, the morphology of inspiral signals can be exploited to offset some
of the computational complexity of known low-latency algorithms.  First, the signals
evolve slowly in frequency, so that they can be broken into contiguous
band-limited time intervals and processed at possibly lower sample rates.
Second, inspiral filter banks consist of highly similar templates, admitting
methods such as the singular value decomposition (\SVD)~\citep{Cannon:2010p10398}
or the Gram-Schmidt process~\citep{rbf} to reduce the number of templates.

Several efforts that exploit one or both of these properties are under way to
develop low-latency \CBC\ search pipelines with tractable computing requirements.
One example is \mbta{}~\citep{Marion2004, Buskulic2010}, which was deployed in
S6/VSR3.  MBTA consists of multiple, usually two, template banks for different frequency bands, one which is matched to the early inspiral and the other which is matched to the late inspiral.  An excursion in the output of any filter bank triggers coherent reconstruction of the full matched filtered output.  Final triggers are built from the reconstructed matched filter output.  Another novel approach using networks of parallel, second-order infinite impulse response~(IIR) filters is being explored
by \citet{shaunIIR} and \citet{linqingIIR}.

We will use both properties to demonstrate that a very low latency detection statistic
is possible with current computing resources.  Assuming the other technical
sources of latency can be reduced significantly, this could make it possible to
send prompt alerts to the astronomical community.

The paper is organized as follows.  First, we discuss prospects for
early-warning detection.  Then, we provide an overview of our novel
method for detecting CBC signals near real-time.
We then describe a prototype implementation using open source signal
processing software.  To validate our approach we present a case study
focusing on a particular subset of the \NS--\NS\ parameter space.  We
conclude with some remarks on what remains to prepare for the advanced
detector era.

\section{Prospects for early-warning detection and \EM\ follow-up}

\begin{figure}[t]
\includegraphics[width=\columnwidth]{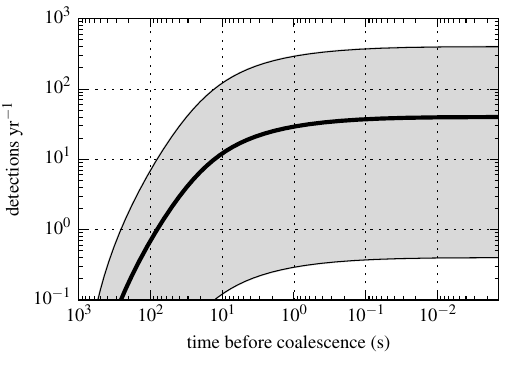}
\caption{\label{fig:earlywarning}Expected number of \NS--\NS\
sources that could be detectable by Advanced \LIGO\ a given number of seconds
before coalescence.  The heavy solid line corresponds to the most probable yearly rate
estimate from~\citet{Abadie:2010p10836}.  The shaded region represents the 5\%--95\% confidence interval
arising from substantial uncertainty in predicted event rates.}
\end{figure}
Before the \GW\ signal leaves the detection band, we can imagine examining the
signal-to-noise ratio (\SNR) accumulated up to that point and if it is
already significant, release an alert immediately, trading \SNR\ and sky
localization accuracy for pre-merger detection.

In the quadrupole approximation, the instantaneous frequency of the GW inspiral signal is related to the time $t$ relative to coalescence ~\citep[section 5.1 of][]{livrev12} through
\begin{equation} \label{eq:fgw}
	f(t) = \frac{1}{\pi \mathcal{M}_\mathrm{t}}
		\left[ \frac{5}{256}\frac{\mathcal{M}_\mathrm{t}}{t} \right]^{3/8},
\end{equation}
where $\mathcal{M}=M^{2/5} \mu^{3/5}$ is the chirp mass of the binary, $\mathcal{M}_\mathrm{t}=G \mathcal{M} / c^3$ is the chirp mass in units of time, $M$ is the total mass, and $\mu$ is the reduced mass. 
The expected value of the single-detector \SNR\ for an optimally oriented (source at detector's zenith or nadir, orbital plane face-on) inspiral source is~\citep{Abadie:2010p10836}
\begin{equation}
	\label{eq:expected-snr}
	\rho =
		\frac{{\mathcal{M}_\mathrm{t}}^{5/6} c}{\pi^{2/3} D}
		\sqrt{
			\frac{5}{6} \int_{f_\mathrm{low}}^{f_\mathrm{high}}
			\frac{f^{-7/3}}{S(f)} \mathrm{d}f},
\end{equation}
where $D$ is the luminosity distance and $S(f)$ is the one-sided power spectral density of the detector noise.  $f_\mathrm{low}$ and $f_\mathrm{high}$ are low- and high- frequency limits of integration which may be chosen to extend across the entire bandwidth of the detector.  If we want to trigger at a time $t$ before merger, then we must cut off the SNR integration at $f_\mathrm{high} = f(t)$ with $f(t)$ given by Equation~(\ref{eq:fgw}) above.

Figure~\ref{fig:earlywarning} shows projected early detectability rates for
\NS--\NS\ binaries in Advanced \LIGO\ assuming the anticipated detector sensitivity for
the `zero detuning, high power' configuration described in \citet{ALIGONoise} and
\NS--\NS\ merger rates estimated in \citet{Abadie:2010p10836}.  The merger rates
have substantial measurement uncertainty due to the small sample of known double pulsar systems that will merge within a Hubble time; they also have systematic uncertainty due to sensitive dependence on the pulsar luminosity distribution function~\citep{KalogeraRates}.
The most probable estimates indicate that at a single-detector \SNR\ threshold of 8, we will
observe a total of 40~events~yr$^{-1}$; $\sim$10~yr$^{-1}$ will be detectable
within 10~s of merger and $\sim$5~yr$^{-1}$ will be detectable within 25~s of
merger if analysis can proceed with near zero latency.

We emphasize that any practical \GW\ search will include technical delays due
to light travel time between the detectors, detector infrastructure, and the
selected data analysis strategy.  Figure~\ref{fig:earlywarning} must be understood
in the context of all of the potential sources of latency, some of which are avoidable
and some of which are not.

\begin{figure}[t]
\includegraphics[width=\columnwidth]{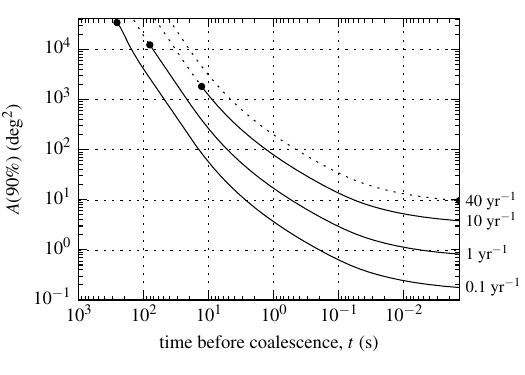}
\caption{\label{fig:sky-localization-accuracy}Area of the 90\% confidence
region as a function of time before coalescence for sources with anticipated
detectability rates of 40, 10, 1, and 0.1~yr$^{-1}$. The heavy dot indicates
the time at which the accumulated \SNR\ exceeds a single-detector threshold of~8.}
\end{figure}
\begin{table}[h]
\caption{\label{table:sky-localization-accuracy}Horizon Distance, \SNR\ at
Merger, and Area of 90\% Confidence at Selected Times Before Merger for Sources
with Expected Detectability Rates of 40, 10, 1, and 0.1~yr$^{-1}$.}
\begin{center}
\begin{tabular}{rrrrrrr}
\tableline\tableline
Rate & Horizon & Final & \multicolumn{4}{c}{$A$(90\%) (deg$^2$)} \\
\cline{4-7}
yr$^{-1}$ & (Mpc) & \SNR\ & 25 s & 10 s & 1 s & 0 s \\
\tableline
40\phd\phn & 445 & 8.0 & ----- & ----- & ----- & 9.6 \\
10\phd\phn & 280 & 12.7 & ----- & 1200 & 78 & 3.8 \\
1\phd\phn & 130 & 27.4 & 1300 & 260 & 17 & 0.8 \\
0.1 & 60 & 58.9 & 280 & 56 & 3.6 & 0.2 \\
\tableline
\end{tabular}
\tablecomments{A dash (-----) signifies that the confidence area\\is omitted because at the indicated time the SNR would\\not have crossed the detection threshold of 8.}
\end{center}
\end{table}

\EM\ follow-up requires estimating the location of the \GW\ source. The localization
uncertainty can be estimated from the uncertainty in the time of arrival of the \GW{}s,
which is determined by the signal's effective bandwidth and \SNR\
\citep{Fairhurst2009}.  Table~\ref{fig:earlywarning} and
Figure~\ref{fig:sky-localization-accuracy} show the estimated 90\%
confidence area versus time of the loudest coalescence events detectable by
Advanced \LIGO\ and Advanced Virgo.  This is the \emph{minimum} area; localization is best at high elevation
from the plane containing the detectors and worst at zero elevation.  \citeauthor{Fairhurst2009} also cautions that his Fisher matrix calculation fails to capture disconnected patches of probability, which occur prominently in networks of three detectors where there are generally two local maxima on opposite sides of the plane of the detectors.  Aside from the mirror degeneracy, characterizing the uncertainty region by the Fisher matrix alone tends to overestimate, rather than underestimate, the area for low-\SNR\ events, but this effect is generally more than compensated by the source being in an unfavorable sky location.  For these reasons, the localization uncertainty estimated from timing is highly optimistic and will only suffice for an order-of-magnitude estimate.  Once per year, we expect to observe an
event with a final single-detector \SNR\ of $\approx$27 whose location can be constrained to about
1300~deg$^2$ (3.1\% of the sky) within 25~s of merger,
260~deg$^2$ (0.63\% of the sky) within 10~s of merger, and
0.82~deg$^2$ (0.0020\% of the sky) at merger.

It is unfeasible to search hundreds of square degrees for a prompt counterpart.  For comparison to some examples of modern ground-based wide-field survey instruments, the Palomar Transient Factory P48 ~\citep{2010SPIE.7735E.122L} has a $3.50 \times 2.31$~deg$^2$ FOV; the Pan-STARRS P1~\citep{2002SPIE.4836..154K} has a 7~deg$^2$ FOV.  Even the eagerly awaited LSST will have an FOV of 9.6~deg$^2$ \citep{2008arXiv0805.2366I}.  However, 
it is possible to reduce the localization uncertainty by only looking at
galaxies from a catalog that lie near the sky location and luminosity distance
estimate from the \GW\ signal~\citep{galaxy-catalog} as was done in S6/VSR3.
Within the expected Advanced \LIGO\ \NS--\NS\ horizon distance,
the number of galaxies that can produce a given signal amplitude is much larger
than in Initial \LIGO\ and thus the catalog will not be as useful
for downselecting pointings for most events. However, exceptional \GW\ sources will
necessarily be extremely nearby. Within this reduced volume there will be fewer
galaxies to consider for a given candidate and catalog completeness will be
less of a concern.  This should reduce the 90\% confidence area substantially.

\section{Novel real-time algorithm for \CBC\ detection}
\label{sec:method}

In this section, we describe a decomposition of the \CBC\ signal space that
reduces \TD\ filtering cost sufficiently to allow for the
possibility of early-warning detection with modest computing requirements.  We
expand on the ideas of \citet{Marion2004} and \citet{Buskulic2010} that describe a
multi-band decomposition of the compact binary signal space that resulted in
a search with minutes latency during S6/VSR3~\citep{HugheyGWPAW2011}.  We combine this
with the \SVD\ rank-reduction method of \citet{Cannon:2010p10398} that exploits
the redundancy of the template banks.

\subsection{Conventional \CBC\ searches}

Searches for inspiral signals typically employ matched filter
banks that discretely sample the possible intrinsic parameters~\citep{findchirppaper}.  Suppose that the observed data $x[k]$ consists of a known, nominal signal $s[k]$, and additive, zero-mean noise $n[k]$
$$
	x[k] = s[k] + n[k].
$$
A matched filter is a linear filter, defined as
$$
	y[k] = \sum_{n=0}^{N-1} h[n] \, x[k-n] = y_s[k] + y_n[k],
$$
where $y_s$ is the response of the filter to the signal alone and $y_n$ is the response of the signal to noise alone.  The matched filter's coefficients maximize the ratio of the expectation of the filter's instantaneous response to the variance in the filter's output:
$$
(\textrm{signal to noise})^2 = \frac{\expectation \left[ y[0] \right]^2}{\var \left[ y[k] \right]} = \frac{y_s[0]^2}{\var \left[ y_n[k] \right]}.
$$
It is well known~\citep[see, for example,][]{matched-filter} that if $n[k]$ is Gaussian and wide-sense stationary, then the optimum is obtained when
$$
\tilde{h}[n] = \tilde{s}^*[n] \, \tilde{S}^{-1}[n],
$$
up to an arbitrary multiplicative constant.  Here, $\tilde{h}[n]$, $\tilde{s}[n]$, and $\tilde{x}[n]$ are the discrete Fourier transforms (DFTs) of $h[k]$, $s[k]$, and $x[k]$, respectively; $\tilde{S}[n] = \expectation \left[ \tilde{n}[n] \tilde{n}^* [n] \right]$ is the folded, two-sided, discrete power spectrum of $n[k]$.  It is related to the continuous, one-sided power spectral density $S(f)$ through
$$
	\tilde{S}[n] =
	\begin{cases}
		S(n) & \textrm{if } n = 0 \textrm{ or } n = N / 2 \\
		S(n f^0 / 2 N) / 2 & \textrm{if } 0 < n < N / 2 \\
		\tilde{S}[N - n] & \textrm{otherwise},
	\end{cases}
$$
where $N$ is the length of the filter and $f^0$ is the sample rate.  (In order to satisfy the Nyquist-Shannon sampling criterion, it is assumed that the detector's continuous noise power spectral density $S(f)$ vanishes for all $f > f^0 / 2$, or alternatively, that the data are low-pass filtered prior to matched filtering.)  The DFT of the output is
\begin{multline}
\label{eq:matched-filter-fd}
\tilde{y}[n] = \tilde{s}^*[n] \, \tilde{S}^{-1}[n] \, \tilde{x}[n] \\
\equiv \left(\tilde{S}^{-1/2}[n] \, \tilde{s}[n]\right)^* \left(\tilde{S}^{-1/2}[n] \, \tilde{x}[n] \right).
\end{multline}
The placement of parentheses in Equation~(\ref{eq:matched-filter-fd}) emphasizes that the matched filter can be thought of as a cross-correlation of a whitened version of the data with a whitened version of the nominal signal.  In this paper, we shall not describe the exact process by which the detector's noise power spectrum is estimated and deconvolved from the data; for the remainder of this paper we shall define $x[k]$ as the \emph{whitened} data stream.  Correspondingly, from this point on we shall use $h[k]$ to describe the \emph{whitened} templates, being the inverse DFT of $\left(\tilde{S}^{-1/2}[n] \, \tilde{s}[n]\right)^*$.

Inspiral signals are continuously parameterized by a set of intrinsic source
parameters $\theta$ that determine the amplitude and phase evolution of the
\GW\ strain. For systems in which the effects of spin can be ignored, the intrinsic
source parameters are just the component masses of the binary,
 $\theta = (m_1, m_2)$. For a given source, the strain observed by the
 detector is a linear combination of two waveforms corresponding to the
`$+$' and `$\times$' \GW\ polarizations.  Thus, we must design two filters
for each $\theta$.

The coefficients for the $\numtmps$ filters are known as templates, 
and are formed by discretizing and time reversing the
waveforms and weighting them by the inverse amplitude spectral density of the
detector's noise.
To construct a template bank, templates are chosen with
$\numtmps/2$ discrete signal parameters $\theta_0,\, \theta_1,\, \dots,\,
\theta_{\numtmps/2-1}$. These are chosen such that any possible signal
will have an inner product $\geqslant$0.97 with at least one template.
Such a template bank is said to have a {\em minimal match} of 0.97~\citep{Owen:1998dk}.

Filtering the detector data involves a convolution of the data with the
templates.  For a unit-normalized template $h_i[k]$ and whitened detector data
$x[k]$, both sampled at a rate $f^0$, the result can be interpreted as the
\SNR, $\rho_i[k]$, defined as
%
%
\begin{equation}
	\label{eq:SNRTD}
	\rho_i [k] = \sum_{n=0}^{N-1} h_{i}[n] \, x [k-n].
\end{equation}
This results in $\numtmps$ \SNR\ time series. Local peak-finding across time and
template indices results in single-detector triggers.  Coincidences are sought
between triggers in different \GW\ detectors in order to form detection candidates.

Equation~(\ref{eq:SNRTD}) can be implemented in the TD as a bank of
finite impulse response (\fir) filters, requiring $\mathcal O(\numtmps
\tmpsamps)$ floating point operations per sample.  However, it is typically
much more computationally efficient to use the convolution theorem and the
fast Fourier transform (\fft) to implement fast convolution in the FD, requiring only
$\mathcal O(\numtmps \lg \tmpsamps)$ operations per sample but incurring
a latency of $\mathcal O(\tmpsamps)$ samples.

\subsection{The \lloid\ method}

Here we describe a method for reducing the computational cost of a \TD\ search
for CBC.  We give a zero latency, real-time algorithm
that competes in terms of floating point operations per second with the
conventional overlap-save \FD\ method, which by contrast requires a significant latency due
to the inherent acausality of the Fourier transform.  Our method, called \lloid\
(Low Latency Online Inspiral Detection), involves two transformations of the
templates that produce a network of orthogonal filters that is far more
computationally efficient than the original bank of matched filters.

The first transformation is to chop the templates into disjointly supported
intervals, or \emph{time slices}.  Since the time slices of a given template
are disjoint in time, they are orthogonal with respect to time.  Given the
chirp-like structure of the templates, the ``early'' (lowest frequency) time
slices have significantly lower bandwidth and can be safely downsampled.
Downsampling reduces the total number of filter coefficients by a factor of
$\sim$100 by treating the earliest part of the waveform at $\sim$$1/100$ of
the full sample rate.  Together, the factor of 100 reduction in the number of
filter coefficients and the factor of 100 reduction in the sample rate during the early inspiral save a
factor of $\sim$$10^4$ floating point operations per second (\flops) over the
original (full sample rate) templates.

However, the resulting filters are still not
orthogonal across the parameter space and are in fact highly redundant.
We use the \SVD\ to approximate the template bank by a set of orthogonal
\emph{basis filters}~\citep{Cannon:2010p10398}.  We find that this approximation
reduces the number of filters needed by another factor of $\sim$100.  These two
transformations combined reduce the number of floating point operations
to a level that is competitive with the conventional high-latency \FD-matched filter approach.  In the remainder of this section we describe the
\lloid\ algorithm in detail and provide some basic computational cost scaling.

\subsubsection{Selectively reducing the sample rate of the data and templates}
\label{sec:time-slices}

The first step of our proposed method is to divide the templates into time
slices in a \TD\ analog to the \FD\ decomposition employed by MBTA~\citep{Marion2004,Buskulic2010}.  The application to GW data analysis is foreshadowed by an earlier \FD\ convolution algorithm, proposed by \citet{gardner1995efficient}, based on splitting the impulse response of a filter into smaller blocks.  We decompose each template
$h_{i}[k]$ into a sum of $S$ non-overlapping templates
\begin{equation}
\label{eq:time-slices}
h_{i}[k] = \sum_{s=0}^{S-1}
	\begin{cases}
		h_i^s[k] & \textrm{if } t^s \leqslant k / f^0 < t^{s+1} \\
		0 & \textrm{otherwise}
	\end{cases}
\end{equation}
for $S$ integers $\{f^0 t^s\}$ such that $0  = f^0 t^0 < f^0 t^1 < \cdots < f^0
t^S = N$.  The outputs of these new time-sliced filters
form an ensemble of partial \SNR\ streams.  By linearity of the filtering
process, these partial \SNR\ streams can be summed to reproduce the
\SNR\ of the full template.

Since waveforms with neighboring intrinsic source parameters $\theta$
 have similar time-frequency evolution, it is possible to design computationally
efficient time slices for an extended region of parameter space rather than to
design different time slices for each template.

For concreteness and simplicity, consider an inspiral waveform in the
quadrupole approximation, for which the time-frequency relation is given by Equation~(\ref{eq:fgw}).
This monotonic time-frequency relationship allows us
to choose time slice boundaries that require substantially less bandwidth at
early times in the inspiral.

An inspiral signal will enter the detection band with some low frequency
$f_\mathrm{low}$ at time $t_\mathrm{low}$ before merger.  Usually the template
is truncated at some prescribed time $t^0$, or equivalent frequency $f_\mathrm{high}$,
often chosen to correspond to the last stable orbit (\LSO). The beginning of the template is critically
sampled at $2 f_\mathrm{low}$, but the end of the template is critically sampled at a
rate of $2 f_\mathrm{high}$. In any time interval smaller than the duration of the template,
the bandwidth of the filters across the entire template bank can be significantly less
than the full sample rate at which data are acquired.

Our goal is to reduce the filtering cost of a
large fraction of the waveform by computing part of the convolution at a lower
sample rate.  Specifically we consider here time slice boundaries with the
smallest power-of-two sample rates that sub-critically sample the time-sliced
templates.  The time slices consist of the $S$ intervals
$\left[t^0, t^1\right),\, \left[t^1, t^2\right),\, \dots,\, \left[t^{S-1}, t^S\right)$,
sampled at frequencies $f^0,\, f^1,\, \dots,\, f^\mathrm{S-1}$, where $f^s$ is at
least twice the highest nonzero frequency component of any filter in the bank for the
$s$th time slice.

The time-sliced templates can then be downsampled in each interval without
aliasing, so we define them as
\begin{equation}
\label{eq:time-sliced-templates}
h_{i}^{s}[k] \equiv
	\begin{cases}
		h_{i}\!\left[k\frac{f}{f^s}\right] & \textrm{if } t^s \leqslant k/f^s < t^{s+1} \\
		0 & \textrm{otherwise.}
	\end{cases}
\end{equation}
We note that the time slice decomposition in Equation~(\ref{eq:time-slices}) is
manifestly orthogonal since the time slices are disjoint in time.  In the next
section, we examine how to reduce the number of filters within each time slice
via \SVD\ of the time-sliced templates.

\subsubsection{Reducing the number of filters with the \SVD}
\label{sec:svd}

As noted previously, the template banks used in inspiral searches are by design
highly correlated.  \citet{Cannon:2010p10398} showed that applying the \SVD\
to inspiral template banks greatly reduces the number of filters required to achieve a
particular minimal match.  A similar technique can be applied to the time-sliced
templates as defined in Equation~(\ref{eq:time-sliced-templates}) above.  The \SVD\
is a matrix factorization that takes the form
\begin{equation}
h_i^s[k] = \sum_{\mathclap{l=0}}^{\mathclap{M-1}} v_{il}^s \sigma_l^s u_l^s[k] \approx \sum_{\mathclap{l=0}}^{\mathclap{L^s-1}} v_{il}^s \sigma_l^s u_l^s[k].
\label{eq:svddecomp}
\end{equation}
where $u_l^s[k]$ are orthonormal \emph{basis templates} related to the original
time-sliced templates through the \emph{reconstruction matrix}, $v_{il}^s\sigma_l^s$.
The expectation value of the fractional loss in \SNR\ is the \SVD\ tolerance, given by
\begin{equation*}
\left[ \sum_{l=0}^{L^s-1} \left( \sigma_l^s \right)^2 \right]\left[ \sum_{l=0}^{M-1} \left( \sigma_l^s \right)^2 \right]^{-1},
\end{equation*}
determined by the number $\numsvdtmps$ of basis templates that are kept in
the approximation.  \citet{Cannon:2010p10398}
showed that highly accurate approximations of inspiral template banks could be
achieved with few basis templates.  We find that when combined with the
time slice decomposition, the number of basis templates \numsvdtmps\ is much
smaller than the original number of templates \numtmps\ and improves on the
rank reduction demonstrated in \citet{Cannon:2010p10398} by nearly an order
of magnitude.

Because the sets of filters from each time slice form orthogonal subspaces, and
the basis filters within a given time slice are mutually orthogonal, the set of
all basis filters from all time slices forms an orthogonal basis spanning the
original templates.

In the next section, we describe how we form our early-warning detection
statistic using the time slice decomposition and the \SVD.

\subsubsection{Early-warning output}

In the previous two sections, we described two transformations that greatly
reduce the computational burden of \TD\ filtering.  We are now prepared to
define our detection statistic, the early-warning output, and to comment on the
computational cost of evaluating it.

First, the sample rate of the detector data must be decimated to match sample
rates with each of the time slices.  We will denote the decimated detector data
streams using a superscript ``$s$'' to indicate the time slices to which they
correspond.  The operator $H^\shortdownarrow$ will represent the appropriate
decimation filter that converts between the base sample rate $f^0$ and the
reduced sample rate $f^s$:
\begin{equation*}
\label{eq:decomp}
	x^{s}[k] = \left( H^\shortdownarrow x^0\right)[k].
\end{equation*}
We shall use the symbol $H^\shortuparrow$ to represent an interpolation filter
that converts between sample rates $f^{s+1}$ and $f^s$ of adjacent time slices,
\begin{equation*}
	x^{s}[k] = \left( H^\shortuparrow x^{s+1}\right)[k].
\end{equation*}

From the combination of the time slice decomposition in
Equation~(\ref{eq:time-sliced-templates}) and the \SVD\ defined in
Equation~(\ref{eq:svddecomp}), we define the early-warning output accumulated
up to time slice $s$ using the recurrence relation,
%
%
\begin{equation}
	\rho_i^s [k] =%
		\overbrace{
			\left(H^\uparrow \rho_i^{s+1}\right)[k]
		}^\textrm{\clap{S/N from previous time slices}}
		+
		\underbrace{
			\sum_{\mathclap{l=0}}^{\mathclap{L^s-1}} v_{il}^s \sigma_l^s
		}_\textrm{\clap{reconstruction}}
		\overbrace{
			\sum_{\mathclap{n=0}}^{\mathclap{N^s-1}} u_l^s[n] x^s[k-n]
		}^\textrm{\clap{orthogonal {\sc fir} filters}} .
\end{equation}
Observe that the early-warning output for time slice 0, $\rho_i^0[k]$,
approximates the \SNR\ of the original templates.
\begin{figure*}[h!]
	\begin{center}
		\includegraphics{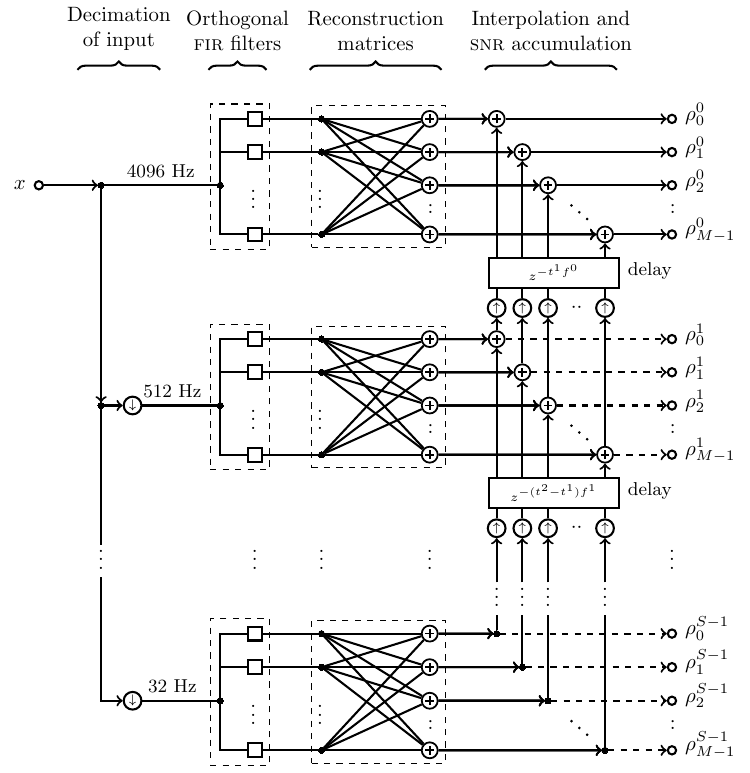}
		\caption{\label{fig:pipeline} Schematic of \lloid\ pipeline illustrating
signal flow.  Circles with arrows represent interpolation
\protect\includegraphics{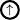} or decimation
\protect\includegraphics{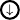}.  Circles with plus
signs represent summing junctions
\protect\includegraphics{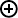}.  Squares
\protect\includegraphics{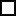} stand for \fir\ filters.  Sample
rate decreases from the top of the diagram to the bottom.  In this diagram, each
time slice contains three \fir\ filters that are linearly combined to produce
four output channels.  In a typical pipeline, the number of \fir\ filters is
much less than the number of output channels.}
	\end{center}
\end{figure*}
The signal flow diagram in Figure~\ref{fig:pipeline} illustrates this
recursion relation as a multirate filter network with a number of early-warning outputs.

Ultimately, the latency of the entire \lloid\ algorithm is set by the decimation
and interpolation filters because they are generally time symmetric and slightly
acausal.  Fortunately, as long as the latency introduced by the decimation and interpolation
filters for any time slice $s$ is less than that time slice's delay $t^s$, the
total latency of the \lloid\ algorithm will be zero.  To be concrete, suppose that the first time slice, sampled at a rate $f^0 = 4096$~Hz, spans times $[t^0,\textrm{ }t^1) = [0\textrm{ s},\textrm{ }0.5\textrm{ s})$, and the second time slice, sampled at $f^1 = 512$~Hz, spans $[t^1,\textrm{ }t^2) = [0.5\textrm{ s},\textrm{ }4.5\textrm{ s})$.  Then the second time slice's output, $\rho_i^1[k]$, will \emph{lead} the first time slice's output, $\rho_i^0[k]$, by 0.5~s.  A decimation filter will be necessary to convert the 4096~Hz input signal $x[k] \equiv x^0[k]$ to the 512~Hz input $x^1[k]$, and an interpolation filter will be necessary to match the sample rates of the two early-warning outputs.  In this example, as long as the decimation and interpolation filters are together acausal by less than $t^1 = 0.5$~s, the total \SNR\ $\rho_i^0[k]$ will be available with a latency of zero samples.  When zero latency is important, we may take this as a requirement for the decimation and interpolation filter kernels.

In the next section, we compute the expected computational cost scaling of this
decomposition and compare it with the direct \TD\ implementation of
Equation~\eqref{eq:SNRTD} and higher latency blockwise \FD\ methods.

\subsection{Comparison of computational costs}

We now examine the computational cost scaling of the conventional \TD\ or
\FD\ matched filter procedure as compared with \lloid.  For convenience,
Table~\ref{tab:recap} provides a review of the notation that we will need in
this section.
%
%
\begin{table}
\caption{\label{tab:recap}Notation Used to Describe Filters.}
\begin{center}
\begin{tabular}{ll}
\tableline\tableline
& Definition \\
\tableline
$f^s$		& Sample rate in time slice $s$ \\
\numtmps		& Number of templates \\
\tmpsamps	& Number of samples per template \\
\numslices	& Number of time slices \\
\numsvdtmps	& Number of basis templates in time slice $s$ \\
\slicessamps	& Number of samples in decimated time slice $s$\\
$N^\shortdownarrow$ & Length of decimation filter \\
$N^\shortuparrow$ & Length of interpolation filter \\
\tableline
\end{tabular}
\end{center}
\end{table}

\subsubsection{Conventional \TD\ method}

The conventional, direct \TD\ method consists of a bank of \fir\ filters, or
sliding-window dot products.  If there are $\numtmps$ templates, each
$\tmpsamps$ samples in length, then each filter requires $M N$ multiplications
and additions per sample, or, at a sample rate $f^0$,
\begin{equation}
	\label{eq:td-flops}
	2 \numtmps \tmpsamps f^0 \textrm{ \flops}.
\end{equation}

\subsubsection{Conventional \FD\ method}

The most common \FD\ method is known as the overlap-save algorithm, described in
\citet{numerical-recipes-chapter-13}.  It entails splitting the input into blocks of $D$
samples, $D > \tmpsamps$, each block overlapping the previous one by $D - \tmpsamps$
samples.  For each block, the algorithm computes the forward \fft\ of the data and
each of the templates, multiplies them, and then computes the reverse \fft.

Modern implementations of the \fft, such as the ubiquitous \texttt{fftw}, require about
$2 \fftblock \lg \fftblock$ operations to evaluate a real transform of size
$\fftblock$~\citep{Johnson:2007p9654}.  Including the forward transform of the data and
$M$ reverse transforms for each of the templates, the \fft\ costs $2 (\numtmps + 1)
\fftblock \lg \fftblock$ operations per block.  The multiplication of the transforms adds
a further $2 \numtmps \fftblock$ operations per block.  Since each block produces
$\fftblock - \tmpsamps$ usable samples of output, the overlap-save method requires
\begin{equation}
	\label{eq:fd-flops}
	f^0 \cdot \frac{2 (\numtmps + 1) \lg \fftblock + 2 \numtmps}{1 - \tmpsamps/\fftblock} \textrm{ \flops}.
\end{equation}

In the limit of many templates, $M \gg 1$, we can neglect the cost of the forward
transform of the data and of the multiplication of the transforms.  The computational
cost will reach an optimum at some large but finite \fft\ block size
$\fftblock \gg \tmpsamps$.  In this limit, the \FD\ method costs
$\approx 2 f^0 \numtmps \lg \fftblock$ \flops.

By adjusting the \fft\ block size, it is possible to achieve low latency with FD
convolution, but the computational cost grows rapidly as the latency in samples
$(D-N$) decreases.  It is easy to show that in the limit of many templates and
long templates, $M, \lg N \gg 1$, the computational cost scales as
$$
\left(1 + \frac{\textrm{template length}}{\textrm{latency}}\right) \left( 2 f^0 M \lg N \right).
$$

\subsubsection{\label{sec:lloid-method}\lloid\ method}

For time slice $s$, the \lloid\ method requires $2 \slicessamps \numsvdtmps f^s$ \flops\ 
to evaluate the orthogonal filters, $2 \numtmps \numsvdtmps f^s$ \flops\ to
apply the  linear transformation from the $\numsvdtmps$ basis templates to the
$\numtmps$ time-sliced templates, and $\numtmps f^s$ \flops\ to add the
resultant partial \SNR\ stream.

The computational cost of the decimation of the detector data is a little bit
more subtle.  Decimation is achieved by applying an \fir\ anti-aliasing filter
and then downsampling, or deleting samples in order to reduce the sample rate
from $f^{s-1}$ to $f^s$.  Naively, an anti-aliasing filter with
$(f^{s-1} / f^s) N^\shortdownarrow$ coefficients should demand
$2 N^\shortdownarrow (f^{s-1})^2 / f^s$ \flops.  However, it is necessary to
evaluate the anti-aliasing filter only for the fraction $f^s / f^{s-1}$ of the
samples that will not be deleted.  Consequently, an efficient decimator 
requires only $2 N^\shortdownarrow f^{s-1}$ \flops.  (One common realization is an ingenious structure called a \emph{polyphase decimator}, described in Chapter 1 of \citet{jovanovic2002multirate}.)

The story is similar for the interpolation filters used to match the sample
rates of the partial \SNR\ streams.  Interpolation of a data stream from a
sample rate $f^s$ to $f^{s-1}$ consists of inserting zeros between the samples
of the original stream, and then applying a low-pass filter with
$(f^{s-1} / f^s) N^\shortuparrow$ coefficients.  The low-pass filter requires
$2 M N^\shortuparrow (f^{s-1})^2 / f^s$ \flops.  However, by taking advantage
of the fact that by construction a fraction $f^s / f^{s-1}$ of the samples are
zero, it is possible to build an efficient interpolator that requires only
$2 M N^\shortuparrow f^{s-1}$ \flops.  (Again, see \citet{jovanovic2002multirate} for a discussion of \emph{polyphase interpolation}.)

Taking into account the decimation of the detector data, the orthogonal \fir\
filters, the reconstruction of the time-sliced templates, the interpolation of
\SNR\ from previous time slices, and the accumulation of \SNR, in total the
\lloid\ algorithm requires
\begin{equation}
\label{eq:lloid-flops}
\sum_{\mathclap{s=0}}^{\mathclap{S-1}} \left( 2 \slicessamps \numsvdtmps + 2 \numtmps \numsvdtmps + \numtmps \right) f^s + 2\sum_{\mathclap{f^s \in \{f^k \, : \, 0 < k < S\}}} \left( N^\shortdownarrow f^0 + \numtmps N^\shortuparrow f^{s-1}\right)
\end{equation}
\flops.  The second sum is carried out over the set of distinct sample rates
(except for the base sample rate) rather than over the time slices themselves,
as we have found that it is sometimes desirable to place multiple adjacent time
slices at the same sample rate in order to keep the size of matrices that enter
the SVD manageable.  Here we have assumed that the
decimation filters are connected in parallel, converting from the base sample
rate $f^0$ to each of the time slice sample rates $f^1$, $f^2$, $\dots$, and
that the interpolation filters are connected in cascade fashion with each
interpolation filter stepping from the sample rate of one time slice to the
next.

We can simplify this expression quite a bit by taking some limits that arise
from sensible filter design.  In the limit of many templates, the cost of the
decimation filters is negligible as compared to the cost of the interpolation
filters.  Typically, we will design the interpolation filters with
$N^\shortuparrow \lesssim \numsvdtmps$ so that the interpolation cost itself is
negligible compared with the reconstruction cost.  Finally, if the number of
basis templates per time slices $\numsvdtmps$ is not too small, the
reconstruction cost dominates over the cost of accumulating the partial \SNR.
In these limits, the cost of \lloid\ is dominated by the basis filters
themselves and the reconstruction, totaling
$2 \sum_{s=0}^{S-1} f^s \numsvdtmps \left( \slicessamps + \numtmps \right)$ \flops.

\subsubsection{Speedup of \lloid\ relative to \TD\ method}

If the cost of the basis filters dominates, and the frequency of the templates
evolves slowly enough in time, then we can use the time-frequency relationship
of Equation~(\ref{eq:fgw}) to estimate the speedup relative to the conventional, direct
\TD\ method.  The reduction in \flops\ is approximately
\begin{multline}
\label{eq:speedup}
\frac{2 \sum_{s=0}^{S-1} f^s \numsvdtmps \slicessamps}{2 \numtmps \tmpsamps f^0} \\
\approx \frac{\alpha}{\left(t_\mathrm{low} - t_\mathrm{high}\right) \left(f^0\right)^2} \int_{t_\mathrm{low}}^{t_\mathrm{high}} \left(2 f(t) \right)^2 \, \mathrm{d} t \\
= \frac{16 \alpha \left(t_\mathrm{low} f^2 (t_\mathrm{low}) - t_\mathrm{high} f^2 (t_\mathrm{high}) \right)}{\left(f^0\right)^2 \left(t_\mathrm{low} - t_\mathrm{high}\right)}
\end{multline}
where $\alpha \approx \numsvdtmps / \numtmps$ is the rank reduction factor, or
ratio between the number of basis templates and the number of templates.  This
approximation assumes that the frequency of the signal is evolving very slowly
so that we can approximate the time slice sample rate as twice the instantaneous
GW frequency, $f^s \approx 2 f(t)$, and the number of samples in
the decimated time slice as the sample rate times an infinitesimally short time
interval, $\slicessamps \approx 2 f(t) \, \mathrm{d}t$. The integral is
evaluated using the power-law form of $f(t)$ from Equation~(\ref{eq:fgw}).
Substituting approximate values for a template bank designed for component
masses around (1.4, 1.4) $M_\odot$, $\alpha \approx 10^{-2}$, $t_\mathrm{low} = 10^3$~s, $f_\mathrm{low} = 10^1$~Hz, $f_\mathrm{high} = f_\mathrm{ISCO} \approx 1570$~Hz, $f^0 = 2 f_\mathrm{ISCO}$, and $t_\mathrm{high} = {f_\mathrm{ISCO}}^{-1}$, we find from Equation~(\ref{eq:speedup}) that the
\lloid\ method requires only $\sim 10^{-6}$ times as many \flops\ as the
conventional \TD\ method.

\section{Implementation}

In this section we describe an implementation of the \lloid\ method described
in Section \ref{sec:method} suitable for rapid \GW\
searches for \CBC{}s.  The \lloid\ method requires several
computations that can be completed before the analysis is underway.  Thus, 
we divide the procedure into an offline planning stage and an
online, low-latency filtering stage.  The offline stage can be done before the
analysis is started and updated asynchronously, whereas the online stage must
keep up with the detector output and produce search results as rapidly as
possible.  In the next two subsections we describe what these stages entail.

\subsection{Planning stage}

The planning stage begins with choosing templates that cover the space of
source parameters with a hexagonal grid~\citep{PhysRevD.76.102004} in order to
satisfy a minimal match criterion.  This assures a prescribed maximum loss in
\SNR\ for signals whose parameters do not lie on the hexagonal grid.  Next, the
grid is partitioned into groups of neighbors called \emph{sub-banks} that
are appropriately sized so that each sub-bank can be efficiently handled by a
single computer.  Each sub-bank contains templates of comparable chirp mass, and
therefore similar time-frequency evolution.  Dividing the source
parameter space into smaller sub-banks also reduces the offline cost of the
\SVD\ and is the approach considered in \citet{Cannon:2010p10398}.  Next, we choose
time slice boundaries as in Equation~\eqref{eq:time-sliced-templates} such that all
of the templates within a sub-bank are sub-critically sampled at progressively lower
sample rates.  For each time slice, the templates are downsampled to the
appropriate sample rate.  Finally, the \SVD\ is applied to each time slice in
the sub-bank in order to produce a set of orthonormal basis templates and a
reconstruction matrix that maps them back to the original templates as
described in Equation~\eqref{eq:svddecomp}.  The downsampled basis templates,
the reconstruction matrix, and the time slice boundaries are all saved to disk.

\subsection{Filtering stage}

The \lloid\ algorithm is amenable to latency-free, real-time implementation.  However, a real-time search pipeline would require integration directly into the data acquisition and storage systems of the \LIGO\ observatories.  A slightly more 
modest goal is to leverage existing low latency, but not real-time, signal processing software in order to implement
the \lloid\ algorithm.

We have implemented a prototype of the low-latency filtering stage using an
open-source signal processing environment called
\gstreamer\footnote{\url{http://gstreamer.net/}} (version 0.10.33).
\gstreamer\ is a vital component of many Linux systems, providing media
playback, authoring, and streaming on devices from cell phones to desktop
computers to streaming media servers.  Given the similarities of
\GW\ detector data to audio data it is not surprising that
\gstreamer\ is useful for our purpose. \gstreamer\ also provides some useful
stock signal processing elements such as resamplers and filters.  We have
extended the \gstreamer\ framework by developing a library called
\gstlal\footnote{\url{https://www.lsc-group.phys.uwm.edu/daswg/projects/gstlal.html}}
that provides elements for \GW\ data analysis.

\gstreamer\ pipelines typically operate with very low (in some consumer
applications, imperceptibly low) latency rather than in true real time because
signals are partitioned into blocks of samples, or \emph{buffers}.  This affords
a number of advantages, including amortizing the overhead of passing signals
between elements and grouping together sequences of similar operations.
However, buffering a signal incurs a latency of up to one buffer length.  This
latency can be made small at the cost of some additional overhead by making the
buffers sufficiently small.  In any case, buffering is a reasonable strategy for low-latency \LIGO\ data analysis because, as we previously remarked, the \LIGO\ data acquisition system has a granularity of $1/16$~s.

\section{Results}

In this section we evaluate the accuracy of the \lloid\ algorithm using our 
\gstreamer{}-based implementation described in the previous section. We calculate
the measured \SNR\ loss due to the approximations of the \lloid\ method and our
implementation of it. Using a configuration that gives acceptable \SNR\ loss
for our chosen set of source parameters, we then compare the computational cost
in \flops\ for the direct \TD\ method, the overlap-save \FD\ method, and \lloid.

\subsection{Setup}
\label{sec:bank-setup}

\begin{figure}[h]
	\includegraphics[width=\columnwidth]{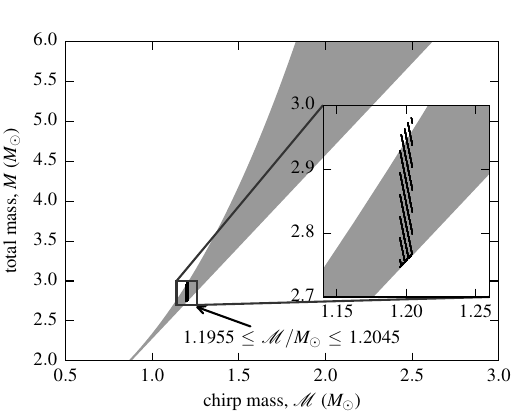}
	\caption{\label{fig:tmpltbank}Source parameters selected for sub-bank used in this
case study, consisting of component masses $m_1$ and $m_2$, between 1 and 3~$M_\odot$, and
chirp masses $\mathcal{M}$ between 1.1955 and 1.2045~$M_\odot$.}
\end{figure}

We examine the performance of the \lloid\ algorithm on a small region of
compact binary parameter space centered on typical \NS--\NS\
masses.  We begin by constructing a template bank that spans component masses
from 1~to~3~$M_\odot$ using a simulated Advanced \LIGO\ noise
power spectrum~\citep{ALIGONoise}\footnote{\url{http://dcc.ligo.org/cgi-bin/DocDB/ShowDocument?docid=T0900288&version=3}}.  Waveforms are generated in the frequency domain in the stationary phase approximation at (post)$^{3.5}$-Newtonian order in phase and Newtonian order in amplitude \citep[the TaylorF2 waveforms described in][]{TaylorF2}.  Templates are truncated at 10~Hz, where the projected
sensitivity of Advanced \LIGO\ is interrupted by the ``seismic wall.''
This results in a grid of 98,544 points, or
$2 \times 98,544 = 197,088$~templates.  Then we create sub-banks by partitioning
the parameter space by chirp mass.  Figure \ref{fig:tmpltbank} illustrates this
procedure. We concentrate on a sub-bank with 657 points with chirp masses
between 1.1955 and 1.2045~$M_\odot$, or $2 \times 657 = 1314$~templates.
With this sub-bank we are able to construct an efficient time slice decomposition
that consists of 11 time slices with sample rates between 32 and 4096 Hz summarized
in Table~\ref{tab:time_slices}.
\begin{table*}
\caption{\label{tab:time_slices} Filter Design Sub-Bank of 1314 Templates.}
\begin{center}
\begin{tabular}{crr@{,\,}lc*{6}{r}}
\tableline\tableline
\\ [-1.5ex]
& $f^s$ & $[t^s$&$t^{s+1})$ & &\multicolumn{6}{c}{$-\log_{10}$ (1$-$\SVD\ tolerance)} \\
\cline{6-11}
\\[-2ex]
& (Hz) & \multicolumn{2}{c}{(s)} & $N^s$ & $1$ & $2$ & $3$ & $4$ & $5$ & $6$ \\ \tableline\\
\multirow{11}{*}{\includegraphics{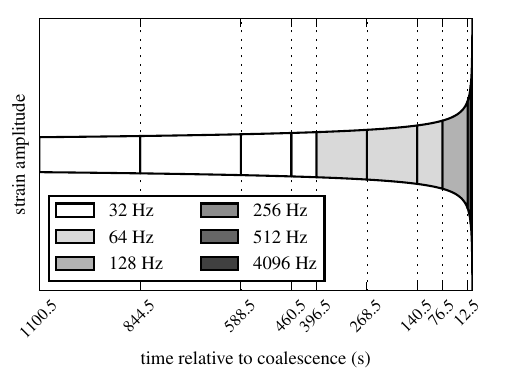}} & 4096 & [0&0.5) & 2048 & 1 & 4 & 6 & 8 & 10 & 14 \\[1em]
& 512 & [0.5&4.5) & 2048 & 2 & 6 & 8 & 10 & 12 & 16 \\[1em]
& 256 & [4.5&12.5) & 2048 & 2 & 6 & 8 & 10 & 12 & 15 \\[1em]
& 128 & [12.5&76.5) & 8192 & 6 & 20 & 25 & 28 & 30 & 32 \\[1em]
& 64 & [76.5&140.5) & 4096 & 1 & 8 & 15 & 18 & 20 & 22 \\[1em]
& 64 & [140.5&268.5) & 8192 & 1 & 7 & 21 & 25 & 28 & 30 \\[1em]
& 64 & [268.5&396.5) & 8192 & 1 & 1 & 15 & 20 & 23 & 25 \\[1em]
& 32 & [396.5&460.5) & 2048 & 1 & 1 & 3 & 9 & 12 & 14 \\[1em]
& 32 & [460.5&588.5) & 4096 & 1 & 1 & 7 & 16 & 18 & 21 \\[1em]
& 32 & [588.5&844.5) & 8192 & 1 & 1 & 8 & 26 & 30 & 33 \\[1em]
& 32 & [844.5&1100.5) & 8192 & 1 & 1 & 1 & 12 & 20 & 23 \\[1em]
\tableline
\end{tabular}
\tablecomments{From
left to right, this table shows the sample rate, time interval, number of
samples, and number of orthogonal templates for each time slice.  We vary \SVD\
tolerance from $\left(1-10^{-1}\right)$ to $\left(1-10^{-6}\right)$.}
\end{center}
\end{table*}
We use this sub-bank and decomposition for the remainder of this section.

\subsection{Measured \SNR\ loss}

The \SNR\ loss is to be compared with the mismatch of 0.03 that arises from the
discreteness of template bank designed for a minimal match of 0.97.  We will consider
an acceptable target \SNR\ loss to be a factor of 10 smaller than this, that is, no more
than 0.003.

We expect two main contributions to the \SNR\ loss to arise in our
implementation of the \lloid\ algorithm.  The first is the \SNR\ loss due to
the truncation of the \SVD\ at $L^s < M$ basis templates.  As remarked upon in
\citet{Cannon:2010p10398} and Section~\ref{sec:svd}, this effect is measured by
the \SVD\ tolerance.  The second comes from the limited bandwidth of the
interpolation filters used to match the sample rates of the partial \SNR\ streams.
The maximum possible bandwidth is determined by the length of the filter,
$N^\shortuparrow$.  \SNR\ loss could also arise if the combination of both the
decimation filters and the interpolation filters reduces their bandwidth
measurably, if the decimation and interpolation filters do not have perfectly uniform
phase response, or if there is an unintended subsample time delay at any stage.

To measure the accuracy of our \gstreamer\ implemention of \lloid\ including all of
the above potential sources of \SNR\ loss, we conducted impulse response tests.  The
\gstreamer\ pipeline was presented with an input consisting of a unit impulse.  By
recording the outputs, we can effectively ``play back'' the templates.  These impulse
responses will be similar, but not identical, to the original, nominal templates.
By taking the inner product between the impulses responses for each output 
channel with the corresponding nominal template, we can gauge exactly how much \SNR\
is lost due to the approximations in the \lloid\ algorithm and any of the technical
imperfections mentioned above.  We call one minus this dot product the \emph{mismatch}
relative to the nominal template.

The two adjustable parameters that affect performance and mismatch the most are
the SVD tolerance and the length of the interpolation filter.  The length of the
decimation filter affects mismatch as well, but has very little impact on
performance.

\paragraph{Effect of \SVD\ tolerance}

\begin{figure*}[b]
	\begin{minipage}[t]{0.5\textwidth}
		\begin{center}
			\includegraphics{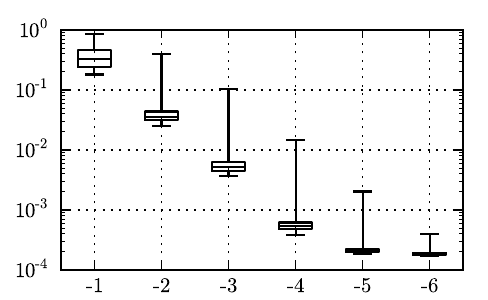}\\
			(a) Mismatch versus \SVD\ tolerance
		\end{center}
	\end{minipage}
	\begin{minipage}[t]{0.5\textwidth}
		\begin{center}
			\includegraphics{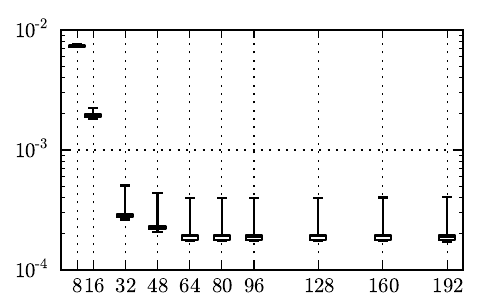}\\
			(b) Mismatch versus $N^\shortuparrow$
		\end{center}
	\end{minipage}
	\caption{\label{fig:bw}Box-and-whisker plot of mismatch between nominal
template bank and \lloid\ measured impulse responses.  The upper and lower boundaries of
the boxes show the upper and lower quartiles; the lines in the center denote the medians.
The whiskers represent the minimum and maximum mismatch over all templates.  In 
(a) the interpolation filter length is held fixed at $N^\shortuparrow = 192$, while
the \SVD\ tolerance is varied from $\left(1-10^{-1}\right)$ to $\left(1-10^{-6}\right)$.
In (b), the \SVD\ tolerance is fixed at $\left(1-10^{-6}\right)$ while $N^\shortuparrow$
is varied from 8 to 192.}
\end{figure*}

We studied how the \SVD\ tolerance affected \SNR\ loss by holding
$N^\shortdownarrow = N^\shortuparrow = 192$ fixed as we varied the \SVD\
tolerance from $\left(1-10^{-1}\right)$ to $\left(1-10^{-6}\right)$.  The
minimum, maximum, and median mismatch are shown as functions of \SVD\ tolerance
in Figure~\ref{fig:bw}(a).  As the \SVD\ tolerance increases toward 1, the \SVD\
becomes an exact matrix factorization, but the computational cost increases as
the number of basis filters increases.  The conditions presented here are more
complicated than in the original work~\citep{Cannon:2010p10398} due to the
inclusion of the time-sliced templates and interpolation, though we still see
that the average mismatch is approximately proportional to the \SVD\ tolerance
down to $\left(1-10^{-4}\right)$.  However, as the \SVD\ tolerance becomes even
higher, the median mismatch seems to saturate around $2 \times 10^{-4}$.  This
could be the effect of the interpolation, or an unintended technical
imperfection that we did not model or expect.  However, this is still an order
of magnitude below our target mismatch of 0.003.  We find that an \SVD\
tolerance of $\left(1-10^{-4}\right)$ is adequate to achieve our target \SNR\
loss.

\paragraph{Effect of interpolation filter length}

Next, keeping the \SVD\ tolerance fixed at $\left(1-10^{-6}\right)$ and the
length of the decimation filter fixed at $N^\shortdownarrow = 192$, we studied
the impact of the length $N^\shortuparrow$ of the interpolation filter on
mismatch.  We use GStreamer's stock
\texttt{audioresample} element, which provides an FIR decimation filter with
a Kaiser-windowed sinc function kernel.  The mismatch as a function of $N^\shortuparrow$ is shown in
Figure~\ref{fig:bw}(b).  The mismatch saturates at $\sim$$2 \times 10^{-4}$ with
$N^\shortuparrow = 64$.  We find that a filter length of 16 is sufficient to
meet our target mismatch of 0.003.

Having selected an SVD tolerance of $\left(1-10^{-4}\right)$ and
$N^\shortuparrow=16$, we found that we could reduce $N^\shortdownarrow$ to 48
without exceeding a median mismatch of~0.003.

We found that careful design of the decimation and interpolation stages made a
crucial difference in terms of computational overhead.  Connecting the
interpolation filters in cascade fashion rather than in parallel resulted in a
significant speedup.  Also, only the shortest interpolation filters that met our
maximum mismatch constraint resulted in a sub-dominant contribution to the
overall cost.  There is possibly further room for optimization beyond minimizing
$N^\shortuparrow$.  We could design custom decimation and interpolation filters,
or we could tune these filters separately for each time slice.

\subsection{Other potential sources of \SNR\ loss}

One possible source of \SNR\ loss for which we have not accounted is the leakage of sharp spectral features in the detector's noise spectrum due to the short durations of the time slices.  In the \lloid\ algorithm, as with many other \GW\ search methods, whitening is treated as an entirely separate data conditioning stage.  In this paper, we assume that the input to the filter bank is already whitened, having been passed through a filter that flattens and normalizes its spectrum.  We elected to omit a detailed description of the whitening procedure since the focus here is on the implementation of a scalable inspiral filter bank.

However, the inspiral templates themselves consist of the GW time series convolved with the impulse response of the whitening filter.  As a consequence, the \lloid\ algorithm must faithfully replicate the effect of the whitening filter.  Since in practice the noise spectra of ground-based \GW\ detectors contain both high-$Q$ lines at mechanical, electronic, and control resonances and a very sharp rolloff at the seismic wall, the frequency response of the \lloid\ filter bank must contain both high-$Q$ notches and a very abrupt high-pass filter.  \fir\ filters with rapidly varying frequency responses tend to have long impulse responses and many coefficients.  Since the \lloid\ basis filters have, by design, short impulse responses and very few coefficients, one might be concerned about spectral leakage contaminating the frequency response of the LLOID filter bank.

The usual statement of the famous Nyquist-Shannon theorem, stated below as Theorem~\ref{thm:nyquist}, has a natural dual, Theorem~\ref{thm:nyquist-dual}, that addresses the frequency resolution that can be achieved with an \fir\ filter of a given length.

\begin{thm}
	\label{thm:nyquist}
	\citep[After][p. 518]{oppenheim1997signals}
	Let $x(t)$ be a band-limited signal with continuous Fourier transform $\tilde{x}(f)$ such that $\tilde{x}(f') = 0 \; \forall \; f' : |f'| > f_M$.  Then, $x(t)$ is uniquely determined by its discrete samples $x(n/f^0)$, $n = 0, \pm 1, \pm 2, \dots$, if $f^0 > 2 f_M$.
\end{thm}

\begin{thm}
	\label{thm:nyquist-dual}
	Let $x(t)$ be a compactly supported signal such that $x(t') = 0 \; \forall \; t' : |t'| > t_M$.  Then its continuous Fourier transform $\tilde{x}(f)$ is uniquely determined by the discrete frequency components $\tilde{x}(n \, \Delta f)$, $n = 0, \pm 1, \pm 2, \dots$, if $\Delta f < 1 / (2 t_M)$.
\end{thm}

Another way of stating Theorem~\ref{thm:nyquist-dual} is that, provided $x(t)$ is nonzero only for $|t| < 1 / (2 \, \Delta f)$, the continuous Fourier transform can be reconstructed at any frequency $f$ from a weighted sum of $\sinc$ functions centered at each of the discrete frequency components, namely,
$$
	\tilde{x}(f) \propto \sum_{n=-\infty}^{\infty} \tilde{x}\left(n \, \Delta f\right) \sinc \left[ \pi (f - n \, \Delta f) / \Delta f \right].
$$
Failure to meet the conditions of this dual of the sampling theorem results in spectral leakage.  For a TD signal to capture spectral features that are the size of the central lobe of the $\sinc$ function, the signal must have a duration greater than $1/\Delta f$.  If the signal $x(t)$ is truncated by sampling it for a shorter duration, then its Fourier transform becomes smeared out; conceptually, power ``leaks'' out into the side lobes of the $\sinc$ functions and washes away sharp spectral features.  In the \GW\ data analysis literature, the synthesis of inspiral matched filters involves a step called \emph{inverse spectrum truncation} \citep[see][Section VII]{findchirppaper} that fixes the number of coefficients based on the desired frequency resolution.

In order to effectively flatten a line in the detector's noise power
spectrum, the timescale of the templates must be at least as long as
the damping time $\tau$ of the line, $\tau = 2 Q / \omega_0$, where $Q$ is the quality
factor of the line and $w_0$ is the central angular frequency.  To put this into the context of the sampling theorem, in order to resolve a notch with a particular $Q$ and $f_0$, an \fir\ filter must achieve a frequency resolution of $\Delta f \gtrsim \pi f_0 / Q$ and therefore its impulse response must last for at least a time $1/\Delta f = Q / \pi f_0$.  For example,
in the S6 detector configuration known as ``Enhanced LIGO,'' the violin modes \citep{ELIGOSusp} had $Q \sim 10^5$ and $\omega_0 \sim (2 \pi) 340$~rad~s$^{-1}$, for a coherence time $\tau \sim 10^2$~s.

In our example template bank, many of the time slices are much shorter
than this.  However, in summation the time slices have the same duration as the full templates themselves, and the full templates are much longer than many
coherence times of the violin mode.  For this reason, we speculate that 
\lloid\ should be just as robust to sharp line features as
traditional \fft-based searches currently employed in the \GW\ field.  Future works must verify this reasonable supposition with numerical experiments, including impulse response studies similar to the ones presented here but with detector noise power spectra containing lines with realistically high quality factors.

There could, in principle, be lines with coherence times many times
longer than the template duration.  For example, the $Q$ of the violin
modes may increase by orders of magnitude in Advanced \LIGO~\citep{ALIGOSusp}.  Also,
there are certainly narrow lines that are non-stationary.  Both of these
cases can be dealt with by preprocessing $h(t)$ with bandstop filters
that attenuates the lines themselves but also conservatively large neighborhoods
around them.  If such bandstops were implemented as an \fir\ filter, they
could be built into the time slices without any difficulty.

Another way to deal with line features with coherence times much longer than the templates would be to entirely `factor' the whitening out of the \lloid\ filter bank.  Any line features could be notched out in the whitening stage with IIR filters, which can achieve infinitely high $Q$ at just second order.  If the detector data were passed through the whitening filter twice, then time-sliced filters need not depend on the detector's noise power spectral density at all.  In such a variation on the \lloid\ method, the basis filters could be calculated from the \emph{weighted} \SVD~\citep[Chapter 3.6]{WeightedSVD, jackson2003user} of the time-sliced templates, using the covariance of the detector noise as a weight matrix.

\subsection{Lower bounds on computational cost and latency compared to other
methods}

We are now prepared to offer the estimated computational cost of filtering this
sub-bank of templates compared to other methods.  We used the results of the
previous subsections to set the \SVD\ tolerance to $\left(1-10^{-4}\right)$,
the interpolation filter length to 16, and the decimation filter length to 48.
Table~\ref{table:flops} shows the computational cost in \flops\ for the sub-bank
we described above.  For the overlap-save \FD\ method, an \fft\ block size of
$\fftblock = 2 \tmpsamps$ is assumed, resulting in a latency of
$\left(\tmpsamps / f^0\right)$ seconds.  Both the \FD\ method and \lloid\ are
five orders of magnitude faster than the conventional, direct \TD\ method.  However, the
\FD\ method has a latency of over half of an hour, whereas the \lloid\ method, with suitable design of the decimation and interpolation filters, has no more latency than the direct \TD\ method.
\begin{table}
\caption{\label{table:flops}Computational Cost in Flop~s$^{-1}$ and Latency in Seconds
of the Direct \TD\ Method, the Overlap-Save \FD\ Method, and \lloid.}
\begin{center}
\begin{tabular}{lllll}
\tableline\tableline
& Flop~s$^{-1}$ & & Flop~s$^{-1}$ & number of \\
Method & (Sub-bank) & Latency (s) & (\NS--\NS) & Machines \\[0.1em]
\tableline
Direct (\TD) & $4.9\times10^{13}$ & 0 & $3.8\times10^{15}$ & $\sim$$3.8\times10^5$ \\
Overlap-save (\FD) & $5.2\times10^8$ & $2\times10^3$ & $5.9\times10^{10}$ & $\sim$$5.9$ \\
\lloid\ (theory) & $6.6\times10^8$ & 0 & $1.1 \times 10^{11}$ & $\sim$$11$ \\
\lloid\ (prototype) & (0.9 cores) & $0.5$ & ------------ & $\gtrsim$$10$ \\
\tableline
\end{tabular}
\end{center}
\tablecomments{Cost is given
for both the sub-bank described in Section~\ref{sec:bank-setup} and a full
1--3~$M_\odot$ \NS--\NS\ search.  The last column gives the approximate number of machines per
detector required for a full Advanced LIGO \NS--\NS\ search.}
\end{table}

\subsection{Extrapolation of computational cost to an Advanced \LIGO\ search}

Table~\ref{table:flops} shows that the \lloid\ method requires $6.6 \times 10^8$
\flops\ to cover a sub-bank comprising 657 out of the total 98,544 mass pairs.
Assuming that other regions of the parameter space have similar computational
scaling, an entire single-detector search for \NS--\NS\ signals in the
1--3~$M_\odot$ component mass range could be implemented with $(98,544/657)\approx150$ times the cost, or $9.9 \times 10^{10}$~\flops.

We computed the computational cost of a full Advanced \LIGO\ \NS--\NS\ search a
second way by dividing the entire 1--3~$M_\odot$ parameter space into sub-banks
of 657 points apiece, performing time slices and SVDs for each sub-bank, and
tabulating the number of floating point operations using
Expression~(\ref{eq:lloid-flops}).  This should be a much more accurate measure
because template length varies over the parameter space.  Lower chirp mass
templates sweep through frequency more slowly and require more computations
while higher chirp mass templates are shorter and require fewer computations.
Despite these subtleties, this estimate gave us $1.1 \times 10^{11}$~\flops,
agreeing with the simple scaling argument above.

Modern (ca. 2011) workstations can achieve peak
computation rates up to $\sim$$10^{11}$~\flops.  In practice, we expect that a
software implementation of \lloid\ will reach average computation rates that are
perhaps a factor 10 less than this, $\sim$$10^{10}$~\flops\ per machine, due to
non-floating point tasks including bookkeeping and thread synchronization.
Given these considerations, we estimate that a full Advanced
\LIGO, single-detector, \NS--\NS\ search with \lloid\ in will require $\sim$$10$ machines.

By comparison, using the conventional \TD\ method to achieve the same latency costs
$4.9 \times 10^{13}$~\flops\ for this particular sub-bank, and so simply scaling up by the factor of $150$ suggests that it would require $7.4 \times 10^{15}$~\flops\
to search the full parameter space.  To account for the varying sample rate and template duration across the parameter space, we can also directly calculate the cost for the full \TD\ method search using expression~(\ref{eq:td-flops}), resulting in $3.8 \times 10^{15}$~\flops, agreeing within an order of magnitude.  This would require~$\gtrsim$$10^5$ current-day machines.  Presently, the \LIGO\ Data Grid%
\footnote{\url{https://www.lsc-group.phys.uwm.edu/lscdatagrid/}} consists of
only $\sim$$10^4$ machines, so direct \TD\ convolution is clearly impractical.

The overlap-save \FD\ method is slightly more efficient than LLOID for this particular sub-bank, requiring $5.2 \times 10^8$~\flops.  The scaling argument projects that a full \FD\ search would require $7.8 \times 10^{10}$~\flops.  The direct calculation from Expression~(\ref{eq:fd-flops}) gives $5.9 \times 10^{10}$~\flops, in order-of-magnitude agreement.  In this application, the conventional \FD\ search is scarcely a factor of two faster than LLOID while gaining only $0.3$\% in \SNR, but only at the price of thousands of seconds of latency.

\subsection{Measured latency and overhead}

Our \gstreamer\ pipeline for measuring impulse responses contained
instrumentation that would not be necessary for an actual search, including
additional interpolation filters to bring the early-warning outputs back to the
full sample rate and additional outputs for recording signals to disk.

We wrote a second, stripped pipeline to evaluate the actual latency and
computational overhead.  We executed this pipeline on one of the submit
machines of the \LIGO-Caltech cluster, a Sun Microsystems Sun
Fire\texttrademark\ X4600~M2 server with eight quad-core 2.7~GHz AMD
Opteron\texttrademark\ 8384 processors.  This test consumed $\sim$90\% of the
capacity of just one out of the 32 cores, maintaining a constant latency of
$\sim$0.5~s.

The measured overhead is consistent to within an order of magnitude with the
lower bound from the \flops\ budget.  Additional overhead is possibly
dominated by thread synchronization.  A carefully optimized \gstreamer\
pipeline or a hand-tuned C implementation of the pipeline might reduce overhead
further.

The 0.5~s latency is probably due to buffering and synchronization.  The latency
might be reduced by carefully tuning buffer lengths at every stage in the
pipeline.  Even without further refinements, our implementation of the \lloid\
algorithm has achieved latencies comparable to the \LIGO\ data acquisition
system itself.  

\section{Conclusions}

We have demonstrated a computationally feasible filtering algorithm for the rapid
and even early-warning detection of \GW{}s emitted during the coalescence
of \NS{}s and stellar-mass black holes.  It is one part of a complicated
analysis and observation strategy that will unfortunately have other sources of
latency.  However, we hope that it will motivate further work to reduce such
technical sources of \GW\ observation latency and encourage the possibility of
even more rapid \EM\ follow-up observations to catch prompt emission in the
advanced detector era.

\CBC\ events may be the progenitors of some short hard \GRB{}s and are expected
to be accompanied by a broad spectrum of \EM\ signals. Rapid alerts to the
wider astronomical community will improve the chances of detecting an \EM\
counterpart in bands from gamma-rays down to radio. In the Advanced LIGO
era, it appears possible to usefully localize a few rare events prior to the
\GRB{}, allowing multi-wavelength observations of prompt emission. More
frequently, low-latency alerts will be released after merger but may still
yield extended X-ray tails and early on-axis afterglows.

The \lloid\ method is as fast as conventional \fft-based, \FD\ convolution but allows for
latency free, real-time operation.  We anticipate requiring $\gtrsim$40
modern multi-core computers to search for binary \NS{}s using
coincident \GW\ data from a four-detector network.  In the future, additional
computational savings could be achieved by conditionally reconstructing the
\SNR\ time series only during times when a composite detection statistic
crosses a threshold~\citep{svd-compdetstat}.  However, the anticipated required
number of computers is well within the current computing capabilities of the
\LIGO\ Data Grid.

We have shown a prototype implementation of the \lloid\ algorithm using
\gstreamer, an open-source signal processing platform.  Although our prototype
already achieves latencies of less than one second, further fine tuning may
reduce the latency even further.  Ultimately the best
possible latency would be achieved by tighter integration between data
acquisition and analysis with dedicated hardware and software. This could be
considered for third-generation detector design.  Also possible for
third-generation instruments, the \lloid\ method could provide the input for a
dynamic tuning of detector response via the signal recycling mirror to match
the frequency of maximum sensitivity to the instantaneous frequency of the
\GW\ waveform.  This is a challenging technique, but it has the potential for
substantial gains in \SNR\ and timing accuracy \citep{PhysRevD.47.2184}.

Although we have demonstrated a computationally feasible statistic
for advance detection, we have not yet explored data calibration and whitening,
triggering, coincidence, and ranking of GW candidates in a
framework that supports early \EM\ follow-up.  One might explore these and also
using the time slice decomposition and the \SVD\ to form low-latency
signal-based vetoes (e.g.,~$\chi^2$~statistics) that have been essential for
glitch rejection used in previous \GW\ \CBC\ searches.  These additional stages
may incur some extra overhead, so computing requirements will likely be somewhat
higher than our estimates.

Future work must more deeply address sky localization accuracy in a
realistic setting as well as observing strategies. Here, we have followed
\citet{Fairhurst2009} in estimating the area of 90\% localization confidence in
terms of timing uncertainties alone, but it would be advantageous to use a
galaxy catalog to inform the telescope tiling \citep{galaxy-catalog}. Because
early detections will arise from nearby sources, the galaxy catalog technique
might be an important ingredient in reducing the fraction of sky that must be
imaged.  Extensive simulation campaigns incorporating realistic binary merger
rates and detector networks will be necessary in order to fully understand the
prospects for early-warning detection, localization, and \EM\ follow-up using
the techniques we have described.

\acknowledgements

\LIGO\ was constructed by the California Institute of Technology and
Massachusetts Institute of Technology with funding from the National Science
Foundation and operates under cooperative agreement PHY-0107417.  C.H. thanks Ilya Mandel for many discussions about rate estimates and the prospects
of early detection, Patrick Brady for countless fruitful conversations about low-latency analysis methods, and John Zweizig for discussions about \LIGO\ data
acquisition.  N.F. thanks Alessandra Corsi and Larry Price for
illuminating discussions on astronomical motivations.  L.S. thanks Shaun
Hooper for productive conversations on signal processing.  This research is
supported by the National Science Foundation through a Graduate Research
Fellowship to L.S. and by the Perimeter Institute for Theoretical Physics through
a fellowship to C.H. D.K. is supported from the Max Planck Gesellschaft.
M.A.F. is supported by NSF Grant PHY-0855494.

This paper has \LIGO\ Document Number {LIGO-P0900004-v34}.

\appendix

\section{\label{sec:low-frequency-cutoff}Low frequency cutoff for inspiral searches}

Ground-based GW detectors are unavoidably affected at low frequencies by seismic and anthropogenic ground motion.  The LIGO test masses are suspended from multiple-stage pendula, which attenuate ground motion down to the pole frequency.  In the detector configuration in place during S6, seismic noise dominated the instrumental background below about 40~Hz.  Considerable effort is being invested in improving seismic attenuation in Advanced LIGO using active and passive isolation~\citep{0264-9381-27-8-084006}, so that suspension thermal noise will dominate down to 10--15~Hz.  Inspiral waveforms are chirps of infinite duration, but since an interferometric detector's noise diverges at this so-called ``seismic wall,'' templates for matched filter searches are truncated at a low-frequency cutoff $f_\mathrm{low}$ in order to save computational overhead with negligible loss of SNR.

\begin{figure*}[b]
	\begin{center}
	\includegraphics{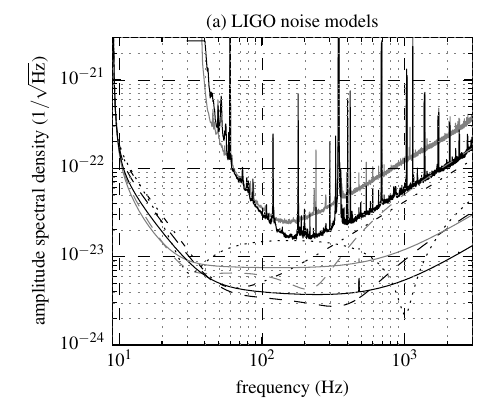}%
	\includegraphics{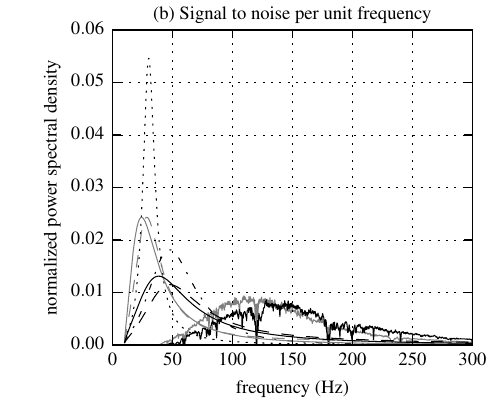}\\[1em]

	\includegraphics{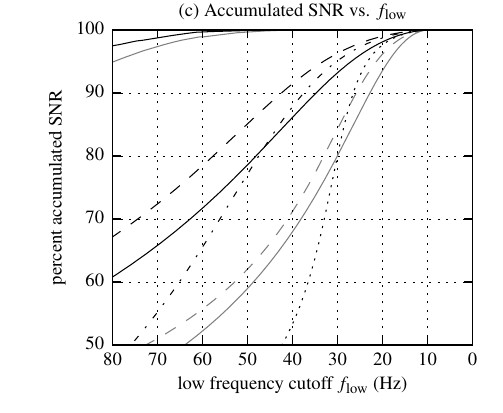}%
	\includegraphics{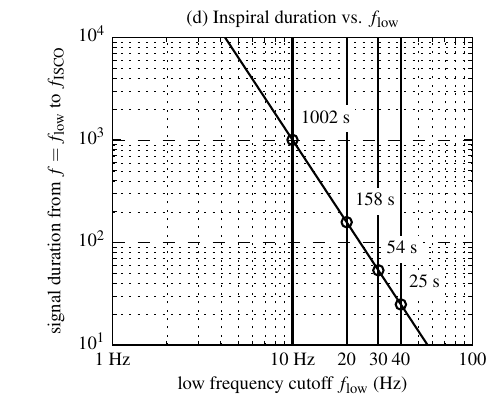}\\[1em]

	\includegraphics{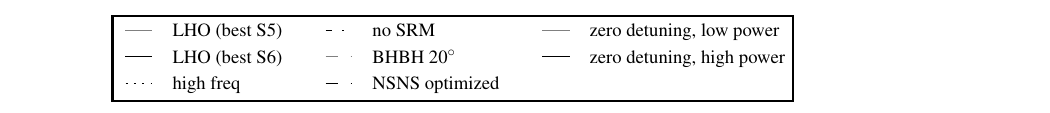}
	\end{center}

	\caption{\label{fig:low-frequency-cutoff}From top left:  (a)~noise amplitude spectral density for a variety of Advanced LIGO noise models, S5, and S6.  (b)~Normalized signal-to-noise per unit frequency, $(\mathrm{d}\rho^2/\mathrm{d}f)/\rho^2$, for a (1.4,~1.4)~$\Msun$ inspiral.  (c)~Percentage of \SNR\ that is accumulated from $f_\mathrm{low}$ to $f_\mathrm{LSO}$, relative to \SNR\ accumulated from $f_\mathrm{low} = 0\,\mathrm{Hz}$ to $f_\mathrm{LSO}$.    (d)~Amount of time for a \NS--\NS\ inspiral signal to evolve from frequency $f_\mathrm{low}$ to $f_\mathrm{LSO}$, as a function of $f_\mathrm{low}$.  For (a)--(c), the line style indicates which noise model was used.}
\end{figure*}

The expected matched-filter SNR, integrated from $f_\mathrm{low}$ to $f_\mathrm{high}$, is given by Equation~(\ref{eq:expected-snr}).  The high-frequency cutoff for the inspiral is frequently taken to be the GW frequency at the LSO; for non-spinning systems, $f_\mathrm{LSO} = 4400 (\Msun / M)$~Hz, where $M$ is the total mass of the binary \citep[section 3.4.1 of][]{livrev12}.  The choice of $f_\mathrm{low}$ is based on the fraction of the total SNR that is accumulated at frequencies above $f_\mathrm{low}$.  To illustrate the relative contributions to the SNR at different frequencies for a (1.4,~1.4)~$\Msun$ binary, we normalized and plotted the integrand of Equation~(\ref{eq:expected-snr}), the noise-weighted power spectral density of the inspiral waveform, in Figure~\ref{fig:low-frequency-cutoff}(b).  This is the quantity
$$
	\frac{1}{\rho^2}\frac{\mathrm{d}\rho^2}{\mathrm{d}f} = \frac{f^{-7/3}}{S(f)} \left( \int_0^{f_\mathrm{LSO}} \frac{{f'}^{-7/3}}{S(f')} \, \mathrm{d}f' \right)^{-1},
$$
which is normalized by the total SNR squared in order to put detectors with different absolute sensitivities on the same footing.  We used several different noise power spectra: all of the envisioned Advanced LIGO configurations from~\citet{ALIGONoise}; the best-achieved sensitivity at LIGO Hanford Observatory (LHO) in LIGO's fifth science run (S5), measured by \citet{S5InspiralRange}; and the best-achieved sensitivity at LHO during S6, measured by \citet{S6InspiralRange}.  (The noise spectra themselves are shown in Figure~\ref{fig:low-frequency-cutoff}(a).)  It is plain that during S5 and S6 the greatest contribution to the \SNR\ was between 100 and 150~Hz, but for all of the proposed Advanced LIGO configurations the bulk of the \SNR\ is accumulated below 60~Hz.  This information is presented in a complementary way in Figure~\ref{fig:low-frequency-cutoff}(c), as the square root of the cumulative integral from $f_\mathrm{low}$ to $f_\mathrm{LSO}$, interpreted as a fraction of the total ``available'' \SNR,
$$
	\rho_\mathrm{frac}(f_\mathrm{low}) = \sqrt{\left( \int_{f_\mathrm{low}}^{f_\mathrm{LSO}} \frac{{f}^{-7/3}}{S(f)} \, \mathrm{d}f \right) \left( \int_0^{f_\mathrm{LSO}} \frac{{f}^{-7/3}}{S(f)} \, \mathrm{d}f \right)^{-1}}.
$$
Table~\ref{table:accum_snr} shows the fractional accumulated \SNR\ for four selected low-frequency cutoffs, 40~Hz, 30~Hz, 20~Hz, and 10~Hz.  In S5 and S6, all of the \SNR\ is accumulated above 40~Hz.  For the `high frequency' Advanced LIGO configuration, scarcely \emph{half} of the \SNR\ is accumulated above 40~Hz.  For the preferred final configuration, `zero detuning, high power,' 86.1\% of the \SNR\ is above 40~Hz, 93.2\% is above 30~Hz, and 98.1\% is above 20~Hz.  (Since \SNR\ accumulates in quadrature, this means, on the other hand, that under the `high frequency' configuration a template encompassing \emph{just the early inspiral} from 10 to 40~Hz would accumulate $\sqrt{1 - 0.533^2} \approx 84.6\%$ of the total \SNR!  In the `zero detuning, high power,' configuration, integration from 10 to 40~Hz alone would yield 50.9\% of the total \SNR, from 10 to 30~Hz, 36.2\%, and from 10 to 20~Hz, 19.4\%.)

\begin{table}[h]
\begin{center}
\caption{\label{table:accum_snr}Fractional Accumulated \SNR\ $\rho_\mathrm{frac}(f_\mathrm{low})$ for Four Selected Low Frequency Cutoffs, $f_\mathrm{low}=40$~Hz, 30~Hz, 20~Hz, and 10~Hz.}
\begin{tabular}{rcccc}
\hline\hline
Noise model & 40 Hz & 30 Hz & 20 Hz & 10 Hz \\
\hline
LHO (best S5) & 100.0 & 100.0 & 100.0 & 100.0 \\
LHO (best S6) & 100.0 & 100.0 & 100.0 & 100.0 \\
High frequency & 53.3 & 80.1 & 97.6 & 100.0 \\
No SRM & 87.8 & 95.1 & 98.7 & 100.0 \\
BHBH 20$^\circ$ & 71.1 & 84.2 & 96.2 & 100.0 \\
NSNS optimized & 91.5 & 96.3 & 99.0 & 100.0 \\
Zero detuning, low power & 67.9 & 80.0 & 93.5 & 100.0 \\
Zero detuning, high power & 86.1 & 93.2 & 98.1 & 100.0 \\
\hline
\end{tabular}
\end{center}
\end{table}

Since the GW amplitude is inversely proportional to the luminosity distance of the source, and the sensitive volume is proportional to distance cubed, the rate of detectable coalescences depends on the choice of low-frequency cutoff.  An inspiral search that is designed with a low-frequency cutoff at the seismic wall would gain an increase in detection rate of $\rho_\mathrm{frac}^{-3}(f_\mathrm{low})$ relative to a search with a low-frequency cutoff of $f_\mathrm{low}$.  This would represent almost a twofold increase in the rate of detection over a search with a fractional accumulated \SNR\ of 80\%, and still a 37\% increase over a search with $\rho_\mathrm{frac} = 90\%$.  Existing coalescing binary detection pipelines strive to sacrifice no more than 3\% of the available \SNR; this forfeits less than a 10\% gain in detection rate.  In order to satisfy this constraint, the low-frequency cutoff would have to be placed below 30~Hz for all of the conceived Advanced LIGO configurations.

The instantaneous GW frequency, given by Equation~(\ref{eq:fgw}), is a power law function of time, so the amount of time for the GW frequency to evolve from $f_\mathrm{low}$ to $f_\mathrm{LSO}$ depends strongly on $f_\mathrm{low}$.  The duration of a (1.4,~1.4)~$\Msun$ inspiral is show in Figure~\ref{fig:low-frequency-cutoff}(d).  The inspiral takes only 25~s to evolve from 40~Hz to $f_\mathrm{LSO}$, but takes 54~s to evolve from 30~Hz to $f_\mathrm{LSO}$, 158~s from 20~Hz, and 1002~s from 10~Hz.

\bibliographystyle{apj}
\bibliography{references}

\end{document}